\documentclass[aps,prc,preprint,superscriptaddress,showpacs]{revtex4}

\newcommand{\beq}{\begin{equation}}
\newcommand{\eeq}{\end{equation}}

\usepackage{epsfig}

\begin{document}
\title{Coupled-channels study of the $\pi^{-}p \to \eta n$ process}

\author{J. Durand }
\affiliation{ Institut de Recherche sur les lois Fondamentales 
de l'Univers, DSM/IRFU, CEA/Saclay, 91191 Gif-sur-Yvette, France}

\author{B. Juli\'a-D\'{\i}az}
\affiliation{ Facultat de Fisica, Universitat de Barcelona,
E-08028 Barcelona, Spain }

\affiliation{ Excited Baryon Analysis Center (EBAC), Thomas Jefferson National
Accelerator Facility, Newport News, VA 22901, USA}

\author{T.-S. H. Lee}

\affiliation{ Excited Baryon Analysis Center (EBAC), Thomas Jefferson National
Accelerator Facility, Newport News, VA 22901, USA}

\affiliation{ Physics Division, Argonne National Laboratory, 
Argonne, IL 60439, USA }

\author{B. Saghai}
\affiliation{ Institut de Recherche sur les lois Fondamentales 
de l'Univers, DSM/IRFU, CEA/Saclay, 91191 Gif-sur-Yvette, France}

\author{T. Sato}
\affiliation{Department of Physics, Osaka University, Toyonaka, 
Osaka 560-0043, Japan}

\affiliation{ Excited Baryon Analysis Center (EBAC), Thomas Jefferson National
Accelerator Facility, Newport News, VA 22901, USA}

\date{\today}
%
%
\begin{abstract}

The reaction $\pi^{-}p \to \eta n$ is investigated within
a dynamical coupled-channels model of
meson production reactions in the nucleon resonance region.
The meson baryon
channels included are $\pi N$, $\eta N$, $\pi \Delta$, $\sigma N$, 
and $\rho N$. The non-resonant meson-baryon
interactions of the model are derived from a set of Lagrangians
by using a unitary transformation method. One or two excited nucleon
states in each of $S$, $P$, $D$, and $F$ partial waves are included
to generate the resonant amplitudes.  Data of $\pi^{-}p \to \eta n$
reaction 
from threshold up to a total center-of-mass energy of about 2 GeV are satisfactorily 
reproduced and the roles played by the following nine nucleon resonances are 
investigated:
$S_{11}(1535)$, $S_{11}(1650)$, $P_{11}(1440)$, $P_{11}(1710)$, $P_{13}(1720)$,
$D_{13}(1520)$, $D_{13}(1700)$, $D_{15}(1675)$, and $F_{15}(1680)$. The reaction
mechanism as well as the predicted $\eta N$ scattering length are discussed.
\end{abstract}

\pacs{ 11.80.-m, 11.80Gw, 13.75.-n, 24.10.Eq } \maketitle

\maketitle
%
%

\section{introduction}\label{Intro}

In spite of the quasi extinction of pion beams facilities since about
two decades, we are witnessing a growing interest in theoretical 
investigations of pion $-$ nucleon ($\pi N$) interactions. 
This is mainly due to the well recognized fact~\cite{Burkert04,Matsuyama07}
 that the impressive amount
of high quality data on electromagnetic meson production reactions from several 
facilities (ELSA, GRAAL, JLab, LEPS, and MAMI) can be used to
pin down the underlying reaction mechanisms and to study the role and/or properties 
of intervening baryon resonances only when the corresponding hadronic production 
reactions can also be consistently understood.
The present work is a prelude to a comprehensive study of the process 
$\gamma p \to \eta p$, 
where, regardless of the direct production mechanisms considered, a meaningful
determination of the resonances properties from the $\eta$ photoproduction data
requires the inclusion of intermediate and final state meson-nucleon interactions.
This latter task is tackled to in the present work by analyzing the
world data of $\pi^- p \to \eta n$ reaction.

To see the main features of our approach, 
it is useful to  briefly describe here some of the  recent theoretical works
which account for the data of $\pi^-p \to \eta n$ reaction.
The K-matrix coupled-channels approach by Sauermann {\it et al.}~\cite{Sauermann95}, 
included only $\pi N$ and $\eta N$ channels and was limited to the $S_{11}$ partial 
wave.
Such a K-matrix approach was then extended by Green and Wycech~\cite{Green99} to 
include the $\gamma N$ channel in a combined analysis of both $\pi^-p \to \eta n$ 
and $\gamma p \to \eta p$ reactions, and more extensively developed
by the Giessen Group~\cite{Feuster98,Penner02a,Penner02b} to  
include $\pi N$, $\eta N$, $2 \pi N$, $\omega N$, $K \Lambda$, and 
$K \Sigma$ channels. 
The approach developed by the Bonn Group~\cite{Anisovich05,Sarantsev05} is also a 
K-matrix coupled-channels model supplemented with Regge phenomenology. 
The approach taken by the Zagreb Group~\cite{Batinic95a,Batinic95b,Batinic97,Ceci06} 
has concentrated on performing the partial-wave analysis of $\pi N \to \pi N,~\eta N$ 
reactions data.
This latter approach is most extensively developed by 
the Virginia Polytechnic Institute - George Washington University (SAID)
Group~\cite{Arndt06}, and is regularly updated.

In this work, we start with a dynamical coupled-channels model 
which is based on a Hamiltonian formulation~\cite{Matsuyama07} and
was applied~\cite{Julia07} to analyze $\pi N$ elastic scattering data. 
This theoretical framework, embodying the $\pi N$, $\eta N$,
$\pi \Delta$, $\sigma N$, and $\rho N$ channels, is an extension of
the approach of Ref.~\cite{Sato96} and is rather 
different from the K-matrix models described above, as discussed in
detail in Refs.~\cite{Matsuyama07,Julia07}.
Qualitatively speaking, the K-matrix approaches, which can be 
derived~\cite{Burkert04} from a dynamical formulation by taking
the on-shell approximation, avoid an explicit treatment of the reaction 
mechanisms in the short range region, where we want to map out the
quark-gluon substructure of the excited states ($N^*$) of the nucleon.
Such a simplification in interpreting the data is also not taken in other dynamical approaches
such as those developed recently in Refs.~\cite{Chen03,Chen07,Gasparyan03} and the
earlier works reviewed in Ref.~\cite{Burkert04}. Besides the approaches mentioned above,
attempts~\cite{Capstick99,Zhong07} to introduce subnucleonic degrees of freedom in 
studying $\pi^{-}p \to \eta n$ reaction are also becoming available.

Moreover, combining the dynamical coupled-channels approach and the constituent quark model 
approach~\cite{Saghai07} to study~\cite{Chiang01,Chiang04,Julia06}
$\gamma p \to K^+ \Lambda$ process proves to be a useful step in deepening our
understanding of the baryon spectroscopy and search for missing nucleon 
resonances~\cite{Capstick00}.

This work follows closely the model (JLMS) developed~\cite{Julia07} in a study
of $\pi N$ elastic scattering. The relevant scattering equations are be described 
in Section II. Section III is devoted to model building procedure and evaluation of 
the data base. 
In the same Section, we present our results for differential and total cross-sections 
of the process $\pi^{-}p \to \eta n$, in the center-of-mass energy 
range $W~\lesssim$ 2 GeV, and discuss the main features of the considered
reaction mechanism.
In Section IV the ingredients of the constructed
model are used to predict the $\eta N$ scattering length, as well
as the total cross-section for the process $\eta p \to \eta p$.
Summary and conclusions are reported in Section V.
%
%

\section{Theoretical framework}\label{Theo}

A detailed description of the coupled-channels formalism can 
be found in Refs.~\cite{Matsuyama07,Julia07}. We outline 
here the main ingredients which are necessary to understand 
the procedure followed in the present work. 

The meson baryon ($MB$) transition amplitudes in each partial wave can be written as, 
\begin{eqnarray}
 T_{MB,M^\prime B^\prime}(E)  &=&  
 t^{NR}_{MB,M^\prime B^\prime}(E)
+ 
 t^R_{MB,M^\prime B^\prime}(E) \,,
\label{eq:tmbmb}
\end{eqnarray}
where,
\begin{eqnarray} 
MB \equiv \pi N,~ \eta N,~ \pi\Delta,~ \rho N,~ \sigma N.
\label{eq:MB}
\end{eqnarray}
The full  
amplitudes $T_{M B,M' B'}(E)$ can be directly used to calculate $M B \to
M' B'$ scattering observables. The non-resonant amplitude 
$t^{NR}_{MB,M^\prime B^\prime}(E)$ in Eq.~(\ref{eq:tmbmb}) is defined by the 
coupled-channels equations,
\begin{eqnarray}
t^{NR}_{MB,M^\prime B^\prime}(E)= V_{MB,M^\prime B^\prime}(E)
+\sum_{M^{\prime\prime}B^{\prime\prime}}
V_{MB,M^{\prime\prime}B^{\prime\prime}}(E) \;
G_{M^{\prime\prime}B^{\prime\prime}}(E)    \;
t^{NR}_{M^{\prime\prime}B^{\prime\prime},M^\prime B^\prime}(E)  \, ,
\label{eq:nr-tmbmb}
\end{eqnarray}
with $G_{M^{\prime\prime}B^{\prime\prime}}(E)$ meson-baryon propagators, and,
\begin{eqnarray}
V_{MB,M^\prime B^\prime}(E)= v_{MB,M^\prime B^\prime}
+Z^{(E)}_{{M}{B},{M}^\prime {B}^\prime}(E)\, . 
\label{eq:veff-mbmb}
\end{eqnarray}
The interactions $v_{MB,M'B'}$ are derived from tree-level 
processes by using a unitary transformation method. They are energy 
independent and free of singularities. On the other hand, 
$Z^{(E)}_{{M}{B},{M}^\prime {B}^\prime}(E)$ is induced by the decays of 
the unstable particles ($\Delta$, $\rho$, $\sigma$) and thus contains 
{\it moving} singularities due to the $\pi\pi N$ cuts. As emphasized  
in Ref.~\cite{Julia07}, we neglect that term at this stage.

The second term in the right-hand-side of Eq.~(\ref{eq:tmbmb}) 
is the resonant term defined by
\begin{eqnarray} 
t^R_{MB,M^\prime B^\prime}(E)= \sum_{N^*_i, N^*_j}
\bar{\Gamma}_{MB \to N^*_i}(E) [D(E)]_{i,j}
\bar{\Gamma}_{N^*_j \to M^\prime B^\prime}(E) \,,
\label{eq:tmbmb-r} 
\end{eqnarray}
with the $N^*$ propagator,
\begin{eqnarray}
[D^{-1}(E)]_{i,j} = (E - M^0_{N^*_i})\delta_{i,j} - \bar{\Sigma}_{i,j}(E)\,,
\label{eq:nstar-g}
\end{eqnarray}
where $M_{N^*}^0$ is the bare mass of the resonant state $N^*$, 
and the self-energies are,
\begin{eqnarray}
\bar{\Sigma}_{i,j}(E)= \sum_{MB}\Gamma_{N^*_i\to MB} G_{MB}(E)
\bar{\Gamma}_{MB \to N^*_j}(E) \,.
\label{eq:nstar-sigma}
\end{eqnarray}
The dressed vertex interactions in Eq.~(\ref{eq:tmbmb-r}) and
Eq.~(\ref{eq:nstar-sigma}) are (defining 
$\Gamma_{MB\to N^*}=\Gamma^\dagger_{N^* \to MB}$),
\begin{eqnarray}
\bar{\Gamma}_{MB \to N^*}(E)  &=&  
{ \Gamma_{MB \to N^*}} + \sum_{M^\prime B^\prime}
t^{NR}_{MB,M^\prime B^\prime}(E) 
G_{M^\prime B^\prime}(E)
\Gamma_{M^\prime B^\prime \to N^*}\,, 
\label{eq:mb-nstar} \\
\bar{\Gamma}_{N^* \to MB}(E)
 &=&  \Gamma_{N^* \to MB} +
\sum_{M^\prime B^\prime} \Gamma_{N^*\to M^\prime B^\prime}
G_{M^\prime B^\prime }(E)t^{NR}_{M^\prime B^\prime,M B}(E) \,. 
\label{eq:nstar-mb}
\end{eqnarray}
The parameterization used for $\Gamma_{N^*,MB}$ is explained in 
Ref.~\cite{Julia07}.

The meson-baryon propagators $G_{MB}$ in the above equations are, 
\begin{eqnarray}
G_{MB}(k,E) = \frac{1}{E-E_M(k)-E_B(k)+ i\epsilon} ,
\end{eqnarray}
for the stable particle channels $MB \equiv \pi N,~ \eta N$, and,
\begin{eqnarray}
G_{MB}(k,E) = \frac{1}{E-E_M(k)-E_B(k)-\Sigma_{MB}(k,E)} , 
\label{eq:rgreen}
\end{eqnarray}
for the unstable particle channels $MB \equiv \pi\Delta,~  \rho N,~  \sigma N$.
The self-energies in Eq.~(\ref{eq:rgreen}) are computed explicitly as defined in 
Ref.~\cite{Julia07}.

To solve the coupled-channels equations, Eq.~(\ref{eq:nr-tmbmb}),
we need to regularize the matrix elements of $v_{MB,M'B'}$. 
We include at each meson-baryon-baryon vertex a form factor of 
the following form:
\beq
F(\vec{k},\Lambda)=\Biggl[\frac {\vec{k}^2}{\vec{k}^2+\Lambda^2} \Biggr]^2 ,
\label{eq:ff}
\eeq
with $\vec{k}$ being the meson momentum. For the meson-meson-meson 
vertex of $v^t$, the form factor in Eq.~(\ref{eq:ff}) is also used with $\vec{k}$ 
being the momentum of the exchanged meson. For the contact term $v^c$, 
we regularize it by $F(\vec{k},\Lambda)F(\vec{k'},\Lambda')$. Here we 
follow Ref.~\cite{Julia07} and use the parameter values determined 
there for all non-resonant terms except the ones explicitly mentioned 
in the following Sections. 

With the non-resonant amplitudes generated from solving Eq.~(\ref{eq:nr-tmbmb}), 
the resonant amplitude $t^R_{MB,M'B'}$ in Eq.~(\ref{eq:tmbmb-r}) then depends 
on the bare mass $M^0_{N^*}$ and the bare $N^*\to MB$ vertex functions.
The vertex functions are parameterized in the following way, 
\begin{eqnarray}
{\Gamma}_{N^*,MB(LS)}(k)
&=& \frac{1}{(2\pi)^{3/2}}\frac{1}{\sqrt{m_N}}C_{N^*,MB(LS)}
\left[\frac{\Lambda_{N^*,MB(LS)}^2}{\Lambda_{N^*,MB(LS)}^2
 + (k- k_R)^2}\right]^{(2+L)}
\left[\frac{k}{m_\pi}\right]^{L} \,.
\label{eq:gmb}
\end{eqnarray}
where $L$ and $S$ are the orbital angular momentum and the total spin
of the $MB$ system, respectively. $C_{N^*,MB(LS)}$ measure the meson-nucleon-$N^*$
coupling strength for a specific $LS$ combination of the $MB$ system and are
taken as free parameters, and $k_R$ are parameters fixed by the $\pi N \to \pi
N$ analysis in Ref.~\cite{Julia07}. 
The above parameterization accounts for the threshold $k^L$ dependence and the 
right power $(2+L)$ such that the integration for calculating the dressed vertex 
Eqs.~(\ref{eq:mb-nstar}) and (\ref{eq:nstar-mb}) is finite. 
%
%
\section{Results and discussion}
\label{Res}

The world data base for the process under investigation embodies 1508 
differential and 98 total cross-sections
~\cite{Mor,Pra,Dei,Ric,Deb,Bro,Cro,Fel,Total} for $1.47~\le~W~\le~2.85$ GeV. 
However, those data presented in 12 papers, thesis, and reports have 
been obtained mainly between 1964 and 1980, except for recent results 
from the Brookhaven National Laboratory and using the Crystal 
Ball detector by Morrison~\cite{Mor} and Prakhov {\it et al.}~\cite{Pra}. 
The quality of data obtained before 1980 has been discussed by 
Clajus and Nefkens~\cite{Ben} and emphasized by George Washington 
University~\cite{Arndt06}, Zagreb~\cite{Batinic95a} and 
Giessen~\cite{Penner02a,Penner02b} Groups, underlying inconsistencies 
among different data sets, because of experimental shortcomings and the 
underestimate of systematic uncertainties. This uncomfortable situation 
has leaded various authors to use a reduced data base. For example, the GWU 
Group~\cite{Arndt06} includes in the data base 257 data points, mainly 
from differential cross-section measurements~\cite{Mor,Pra,Dei,Ric}, 
but also about 50 data points for total 
cross-sections~\cite{Mor,Dei,Ric,Cro,Total}.

In the present work, we concentrate only on the differential cross-sections for 
$W~\lesssim~$ 2 GeV, as summarized in Table~\ref{tab:dxs}. The number of data points
included in the fitted data base in this work (294) will be discussed in 
Sec.~\ref{sec:Fit}.

One of the delicate points in dealing with those data is related to the 
systematic uncertainties ($\delta _{sys}$). For the most recent data by 
Prakhov~{\it et al.}~\cite{Pra}
those uncertainties are given clearly by the authors (6\%).
For old data, we have mainly followed the general 
trend suggested in Ref.~\cite{Ben}, as summarized in the last column of 
Table~\ref{tab:dxs}. Deinet~{\it et al.}~\cite{Dei} report two sources of
systematic uncertainty: 7\% and 9\%, to be added up quadratically, 
giving $\delta _{sys}$ = 11.4\%.
For Richards~{\it et al.}~\cite{Ric}, we have used $\delta _{sys}$ = 10\% for
the lowest energies, 11\% to 14\% for other ones, as given in the original paper.
For Debenham~{\it et al.}~\cite{Deb} and Brown~{\it et al.}~\cite{Bro},
we have followed the conclusion of the Zagreb Group~\cite{Batinic95a,Batinic97}.
In the case of Brown~{\it et al.}~\cite{Bro}, we also have lowered the momentum by
4\%, in lines with Ref.~\cite{Ben}.

\begin{table}[ht]
\begin{tabular}{lccccccccc}
\hline
\hline
  Ref. & Angular range &  $P_\pi$ && $W$   && $N_{dp}$ && $N_{dp}$ used in the& $\delta _{sys}$ \\
       & (Degrees)     &  (GeV/c) && (GeV) &&         &&  the present work &  \\
\hline
Prakhov~{\it et al.}~\cite{Pra} & 23~-~157& 0.687~-~0.747 &&  1.49~-~1.52 && 84 && 70 &  6\% \\
Deinet~{\it et al.}~\cite{Dei} & 32~-~123& 0.718~-~1.050 &&  1.51~-~1.70 && 83 &&  80 &  11\%  \\
Richards~{\it et al.}~\cite{Ric} & 26~-~154& 0.718~-~1.433 &&  1.51~-~1.90 && 70 && 66 &  10\% to 14\% \\
Debenham~{\it et al.}~\cite{Deb} & 162~-~172& 0.697~-~0.999 &&  1.49~-~1.67 && 111 && 27 &  10\% + 0.02 mb \\
Brown~{\it et al.}~\cite{Bro} & 18~-~160& 0.724~-~2.724 &&  1.51~-~2.45 && 379 && 51  &  10\% or 0.01 mb \\
Morrison~\cite{Mor} & 46~-~134 & 0.701~-~0.747 &&  1.50~-~1.52 && 34 && - \\
Crouch~{\it et al.}~\cite{Cro} & 14~-~167& 1.395~-~3.839 &&  1.88~-~2.85 && 731 && - \\
Feltesse~{\it et al.}~\cite{Fel} & 20~-~160 & 0.757 &&  1.53 && 16 && - \\
\hline
\hline
\end{tabular}
\caption{Summary of differential cross-section data for the reaction 
$\pi^- p \to \eta n$. Data sets investigated in the present work are 
singled out in the last two columns, where the number of data points 
($N_{dp}$) per data set used in the fitting procedure is given.}
\protect\label{tab:dxs}
\end{table}
 
Total cross-section data has not been included in our fits due to the 
following reasons: 
i) differential cross-sections are measured by various collaborations in significantly 
different angular ranges with respect to extreme ones (see second column in 
Table~\ref{tab:dxs}),
ii) there is no commonly agreed upon procedure to extract 
total cross-sections from measured angular distributions,
iii) model predictions 
for extreme angles do not in general agree with each other.  

%
%
\subsection{Fitting procedure}\label{sec:Fit}

As mentioned above, in Ref.~\cite{Julia07} the $\pi N\to\pi N$ reaction 
was studied within a coupled-channels formalism, with multi-step processes 
embodying $\pi N$, $\eta N$, $\pi \Delta$, $\sigma N$, and $\rho N$ states.

In that work 175 adjustable parameters were introduced to fit amplitudes
produced by the SAID Group, fitting more than 10000 data points. 
About 30 of those parameters are particularly relevant to the coupled-channels 
mechanisms for the $\pi^- p \to \eta n$ reaction. Accordingly, in the 
present work we use that reduced set of adjustable parameters 
(see Table~\ref{tab:params})  and fix the others to their values as determined in 
Ref.~\cite{Julia07} (cf. Tables III to VII in that paper). 
A total of 294 data points are fitted in the present work 
(see Table~\ref{tab:dxs}). Here, we wish to make a comment on the exclusion
of a few data points in Ref.~\cite{Pra} from the fitted data-base. Actually, 
as mentioned above,
on the one hand, recent data by Prakhov et al.~\cite{Pra} bear much smaller
errors than other data, and on the other hand, the data-base suffers from
some inconsistencies. One of the consequences of this situation is that
a few data points introduce large $\chi^2$s (around 10 or more) thus reducing 
significantly the efficiency of the minimization procedure. The excluded
points concern mainly the two lowest energy sets of Ref.~\cite{Pra}. 

In the following we present our results for two models, as well as those
obtained using the parameters reported in Ref.~\cite{Julia07} (see 
Table~\ref{tab:params} and Figs.~\ref{fig:ds-1}-\ref{fig:ds-2}).

Here, in lines with Ref.~\cite{Julia07}, the following nucleon 
resonances ($N^*$) are considered:
$S_{11}(1535)$, $S_{11}(1650)$, $P_{11}(1440)$, $P_{11}(1710)$, $P_{13}(1720)$,
$D_{13}(1520)$, $D_{13}(1700)$, $D_{15}(1675)$, and $F_{15}(1680)$.

The adjustable parameters for non-resonant terms are the $\eta NN$ coupling 
constant $f_{\eta NN}$ and cut-off $\Lambda_{\eta NN}$. For resonant terms the 
parameters are as follows: $N^*$s bare-masses $M_{0}^{N^*}$, $\eta N N^*$ 
coupling strengths $C_{\eta N N^*}$, and cut-offs $\Lambda_{\eta N N^*}$. 
 
{\squeezetable
\begin{table}[hb!]
\begin{tabular}{llccc}
\hline
\hline
Category & Parameter & Model $A$ &  Model $B$ & Ref.~\cite{Julia07} \\
\hline
Non-resonant $\eta N$ parameters:&  &  &   &  \\
             & $f_{\eta NN}$ & 4.9936 &  4.9999 & 3.8892 \\
             & $\Lambda_{\eta NN}$ & 592.11 &  591.91 & 623.56 \\
Bare mass $M_{0}^{N^*}$:  &  &  &   &  \\
          & $S_{11}(1535)$ & 1809 & 1808 & 1800 \\
          & $S_{11}(1650)$ & 1901 & 1861 & 1880 \\
          & $P_{11}(1440)$ & 1775 & 1784  & 1763 \\
          & $P_{11}(1710)$ & 2019 & 2057  & 2037 \\
          & $P_{13}(1720)$ & 1726 & 1691  & 1711 \\
          & $D_{13}(1520)$ & 1918 & 1919  & 1899 \\
          & $D_{13}(1700)$ & 1971 & 1968  & 1988 \\
          & $D_{15}(1675)$ & 1878 & 1878  & 1898 \\
          & $F_{15}(1680)$ & 2207 & 2207 & 2187 \\
$C_{N^*\to MB(LS)}$:&  &  &   &  \\
 & $C_{\eta N S_{11}(1535)}$  & 8.4269 & 7.8344  & 9.1000 \\
 & $C_{\rho N S_{11}(1535)}$  & {\it 2.0280} & -0.4935  & 2.028 \\
 & $C_{\eta N S_{11}(1650)}$  & 2.0487 & -0.4221  & 0.6000 \\
 & $C_{\rho N S_{11}(1650)}$  & {\it -9.5179} & 2.0000  & -9.5179 \\
 & $C_{\eta N P_{11}(1440)}$  & 1.6321 & 1.6298  & 2.6210 \\
 & $C_{\eta N P_{11}(1710)}$  & 2.4925 & 2.4994  & 3.6611 \\
 & $C_{\eta N P_{13}(1720)}$  & 2.4474 & 2.4997  & -0.9992 \\
 & $C_{\eta N D_{13}(1520)}$  & 0.4440 & 0.4267  & -0.0174 \\
 & $C_{\eta N D_{13}(1700)}$  & -1.8985 & -0.6463  & 0.3570 \\
 & $C_{\eta N D_{15}(1675)}$  & 0.2456 & 0.3437  & -0.0959 \\
 & $C_{\eta N F_{15}(1680)}$  & -0.0446 & -0.0265  & 0.0000 \\
$\Lambda_{N^*\to MB (LS)}$:&  &  &   &  \\
 & $\Lambda_{\eta N S_{11}(1535)}$  & 779.38       & 799.90  & 598.97 \\
 & $\Lambda_{\rho N S_{11}(1535)}$  & {\it 1999.8} &  670.89 & 1999.8 \\
&                                    & {\it 1893.8} &  955.8 & 1893.8 \\
 & $\Lambda_{\eta N S_{11}(1650)}$  & 500.07       & 1999.70  & 500.02 \\
 & $\Lambda_{\rho N S_{11}(1650)}$  & {\it 796.83} &  2000.00 & 796.83 \\
 & $\Lambda_{\eta N P_{11}(1440)}$  & 1766.80       & 1757.40  & 1654.85 \\
 & $\Lambda_{\eta N P_{11}(1710)}$  & 500.08       & 500.00  & 897.84 \\
 & $\Lambda_{\eta N P_{13}(1720)}$  & 631.90       & 649.11  & 500.23 \\
 & $\Lambda_{\eta N D_{13}(1520)}$  & 500.20       & 500.01  & 1918.20 \\
 & $\Lambda_{\eta N D_{13}(1700)}$  & 540.55       & 763.13  & 678.41 \\
 & $\Lambda_{\eta N D_{15}(1675)}$  & 507.64       & 500.00  & 1554.00 \\
 & $\Lambda_{\eta N F_{15}(1680)}$  & 811.72       & 1073.80  & 655.87 \\
\hline
~~~~~~~~~~~~~~~~~~~~$\chi^2_{dp}$ &   & 2.03       & 1.94  & 6.94 \\
\hline
\hline
\end{tabular}
\caption{Parameters for models $A$ and $B$ determined in this work. The last
column gives the values determined in Ref.~\cite{Julia07}.}
\protect\label{tab:params}
\end{table}
}

Model $A$ is obtained by fitting the data base and those 29 adjustable 
parameters (see column 3 in Table~\ref{tab:params}). Model $B$, for 
reasons explained below, has five more adjustable parameters, namely, 
the coupling constants and cut-offs of $\rho N S_{11}$, with $S_{11}$ $\equiv$ 
$S_{11}(1535)$ and $S_{11}(1650)$ for [LS] = [0, 1/2], as well as the 
cut off $\Lambda_{\rho N S_{11}(1535)}$ for [LS] = [2, 3/2]. 
Finally, for comparisons, we reproduce 
in Table~\ref{tab:params} the relevant values reported in
Ref.~\cite{Julia07}. As mentioned above, in that latter work, adjustable
parameters are determined {\it via} the $\pi N \to \pi N$ channels, and
for the $\pi^- p \to \eta n$ the data base embodied
only a few total cross-section data from Refs.~\cite{Pra,Bro}. 
Notice that the five $\rho N S_{11}$ 
parameters in model $A$ (shown in italics in Table~\ref{tab:params})
were not treated as adjustable parameters, and hence,
are identical to those of Ref.~\cite{Julia07}. 
\subsection{Differential and total cross-sections for the process 
$\pi^{-}p \to \eta n$}
\label{ds}
\begin{figure}[thb]
\vspace{35pt}
\begin{center}
\mbox{\epsfig{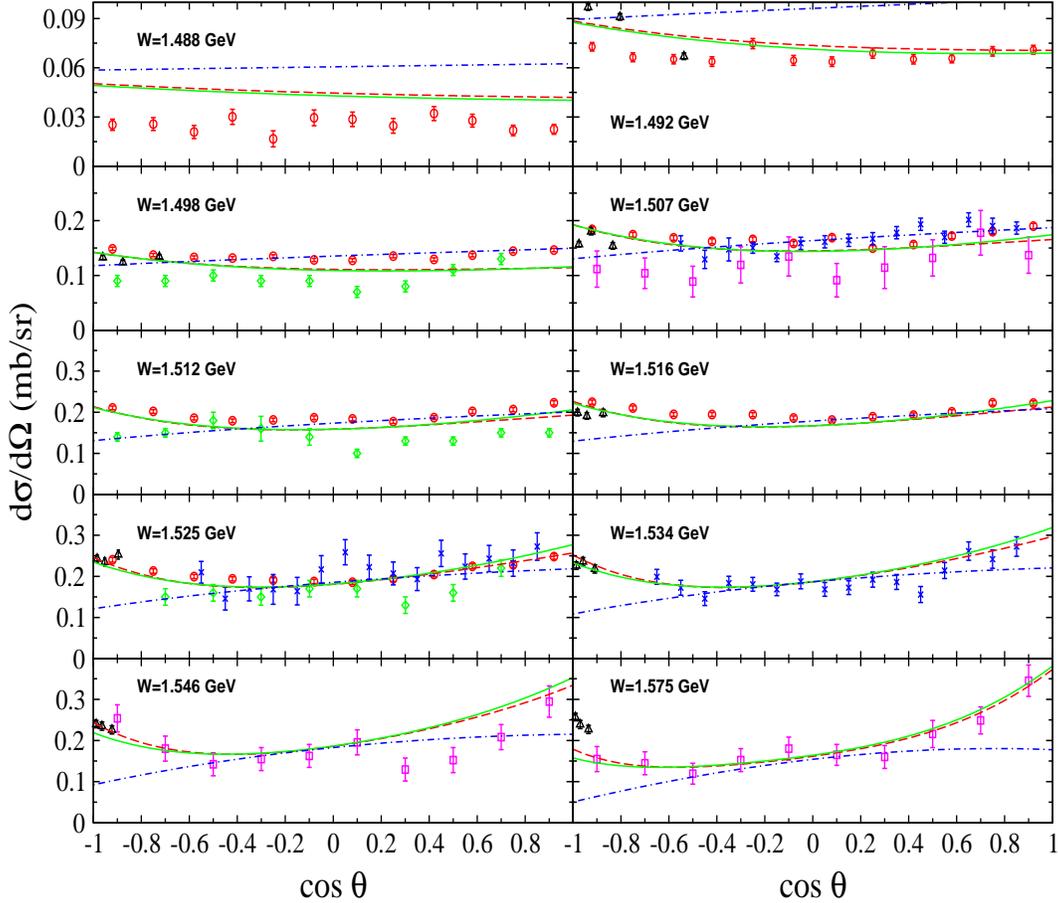}}
\end{center}
\caption{Differential cross-section for the reaction
$\pi^- p \to \eta n$. The curves correspond to models $A$ (dashed) and 
$B$ (full) from the present work. The dash-dotted curves are
obtained by using the parameters in Ref.~\cite{Julia07}.
Data are from Prakhov~{\it et al.}~\cite{Pra} (empty circles), 
Deinet~{\it et al.}~\cite{Dei} (cross),
Richards~{\it et al.}~\cite{Ric} (empty squares),
Debenham~{\it et al.}~\cite{Deb} (up-triangles), and Morrison~\cite{Mor}
(diamonds).
Data uncertainties depicted are only statistic ones.
}
\protect\label{fig:ds-1}
\end{figure}

\begin{figure}[thb]
\vspace{25pt}
\begin{center}
\mbox{\epsfig{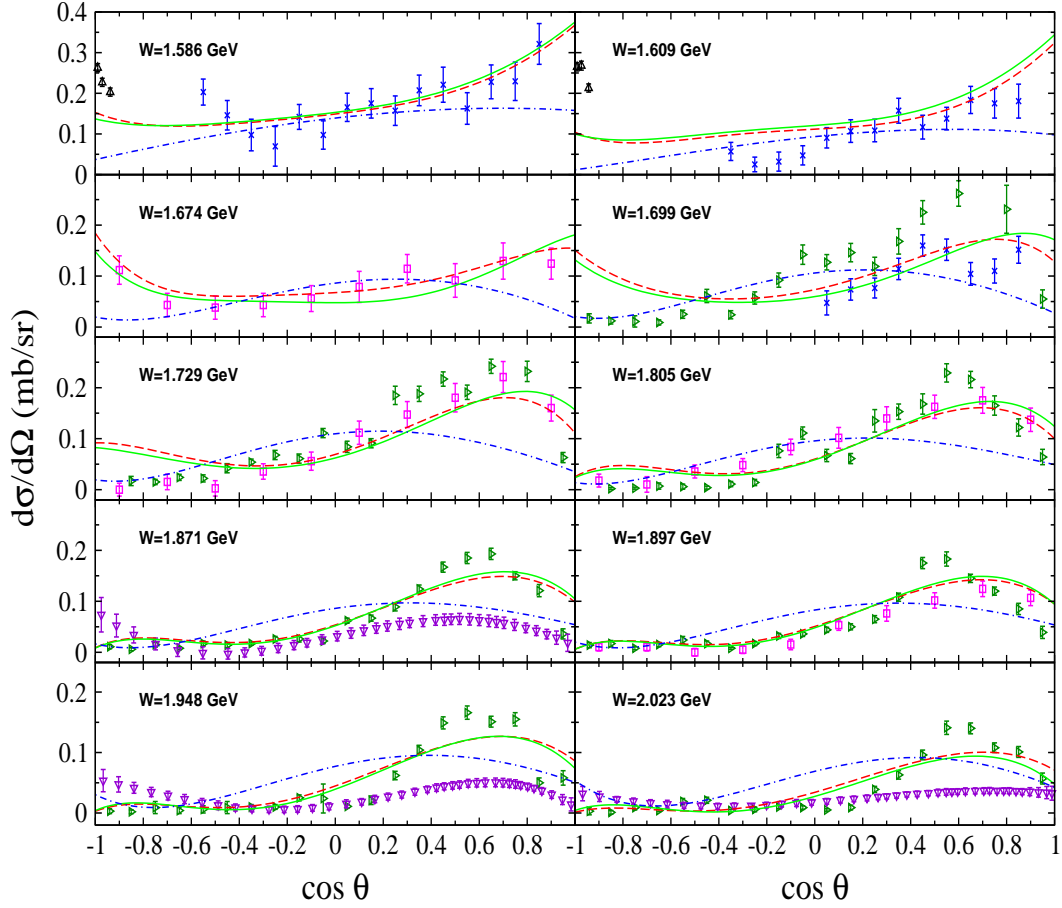}}
\end{center}
\caption{Differential cross-section for the reaction
$\pi^- p \to \eta n$. Curves and data as in Fig.~\ref{fig:ds-1} .
Additional data are from Brown~{\it et al.}~\cite{Bro} (right-triangles)
and Crouch {\it et al.}~\cite{Cro} (down-triangles).
}
\protect\label{fig:ds-2}
\end{figure}

In Figures ~\ref{fig:ds-1} and \ref{fig:ds-2} we compare the results of 
the models $A$ and $B$ with the differential cross-section data, 
for which the reduced $\chi^2$s per data point are 2.03 and 1.94, 
respectively. Those numbers compare well with the GWU~\cite{Arndt06} 
reduced $\chi^2$~=~2.44. In the same figures, we also show results
obtained by using the parameters of Ref.~\cite{Julia07}, which gives
$\chi^2$~=~6.94.

%
%

Before discussing different curves in comparison with data, we wish to 
emphasize the difference between models $A$ and $B$. Once the model 
$A$ obtained, we investigated the importance of various parameters and 
found that by switching off the $\rho$ coupling to the $S_{11}(1535)$, 
the $\chi^2$ increases by roughly a factor of 3. Within the investigated reaction,
such a high sensitivity to the $\rho N S_{11}$ 
seems unrealistic. To cure that behavior, we refitted the data by allowing
those coupling constants to vary in the range of $\pm$0.5 for $S_{11}(1535)$
and $\pm$2 for $S_{11}(1650)$, instead of $\pm$10. 
The model $B$ is then obtained, where that effect is significantly reduced. 
Comparing the two models in Figs.~\ref{fig:ds-1} and \ref{fig:ds-2}, 
we observe that they differ from each other by less than the statistic 
uncertainties of the data, corroborating that the $\pi^- p \to \eta n$
reaction is not a proper channel to pin down those couplings. 
Better constraints on those couplings can be obtained by 
investigating the $\pi N \to \pi \pi N$ process~\cite{juliakamano08}.

%
%

Models $A$ and $B$ show reasonable agreements with Prakhov~{\it et  al.}~\cite{Pra} data, 
except at the lowest energy (Fig.~\ref{fig:ds-1}). 
We will come back to that point later. 
At four common energies, data from Morrison~\cite{Mor} are 
also depicted. That latter data set, not included in our fitting procedure, 
shows systematically smaller cross-sections compared to Ref.~\cite{Pra} data.

Prakhov~{\it et al.}~\cite{Pra} data set at $W~=$~1.507 GeV is of 
special interest, since there are also data from three other 
measurements. Results from Deinet~{\it et al.}~\cite{Dei} are 
compatible with Prakhov~{\it et al.} data, though with larger 
uncertainties (which become even more sizeable at $W~=$~1.525 GeV). 
Richards~{\it et al.}~\cite{Ric} data show deviations from 
Prakhov~{\it et al.} ones, especially at most backward angles. 
Finally, copious data from Debenham~{\it et al.}~\cite{Deb} are 
unfortunately limited to extreme backward angles and appear to 
be rather consistent with other data only up to $W~\approx$~1.55 GeV. 
This trend is confirmed in Fig.~\ref{fig:ds-2}, where models $A$ 
and $B$ reproduce correctly results from Deinet~{\it et al.}~\cite{Dei} 
and Richards~{\it et al.}~\cite{Ric}. Both sets of data come out 
fairly compatible with measurement from Brown~{\it et al.}~\cite{Bro} 
at $W~=$~1.699, 1.729, and 1.805 GeV. Models $A$ and $B$ show 
acceptable agreements with those data, except at backward angles, 
where the model / experiment discrepancies get reduced when 
energy increases and suitable agreement is observed at $W~=$~1.897 GeV. 
At the three remaining depicted energies ($W =$~1.871, 1.948, and 
2.003 GeV) our models reproduce the general trend of 
Brown~{\it et al.} data. At those energies, data from 
Crouch~{\it et al.}~\cite{Cro}, not included in our data base, 
are also shown. The two data sets are not consistent. Given the 
known problems~\cite{Ben} with Brown's data, we made also attempts 
to fit the data base, within model $A$, by replacing the 
Brown~{\it et al.} data by those of Crouch~{\it et al.} at 
$W$~=~1.879 and 1.915 GeV. However, we observed a significant 
increase of $\chi^2$ which goes from 2.03 to 4.12, and with 
very undesirable effects in the Crystal Ball energy range.

%
%
\begin{figure}[htb]
\vspace{35pt}
\begin{center}
\mbox{\epsfig{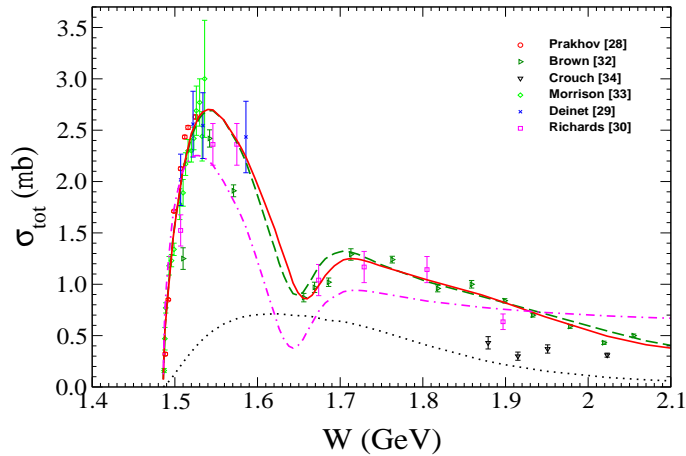}}
\end{center}
\caption{Total cross-section for the reaction
$\pi^- p \to \eta n$. Curves are from Ref.~\cite{Julia07} (dash-dotted),
model $A$ (dashed), model $B$ (full), and the background contributions (dotted) 
in model $B$. 
Data are as in Figs.~\ref{fig:ds-1} and \ref{fig:ds-2}.
}
\protect\label{fig:tot}
\end{figure}

In Figures ~\ref{fig:ds-1} and \ref{fig:ds-2}, results using the 
parameters in Ref.~\cite{Julia07} are also shown. At lowest energies, that model 
overestimates the data. At higher energies, it shows significant 
deviations, first at backward angles and then at forward 
angles. Above $W~\approx$~1.8 GeV it tends to miss the data.

Finally, as mentioned above, we did not include the extracted total 
cross-section data in our data-base. In Fig.~\ref{fig:tot}, 
we show the postdictions of our models $A$ and $B$, as well 
as results of the Ref.~\cite{Julia07}, and compare them with the data.
Both models $A$ and $B$ reproduce correctly the data, except for those
by Crouch {\it et al.}~\cite{Cro}, for which the differential cross-sections
turn out to be significantly smaller than other data, as shown in 
Fig.~\ref{fig:ds-2}. Moreover, the background contributions show a smooth
behavior and are small with respect to the full model results, except
around the minimum of the total cross-section, where resonant terms
produce highly destructive interferences.

%
%
\begin{figure}[htb]
\vspace{35pt}
\begin{center}
\mbox{\epsfig{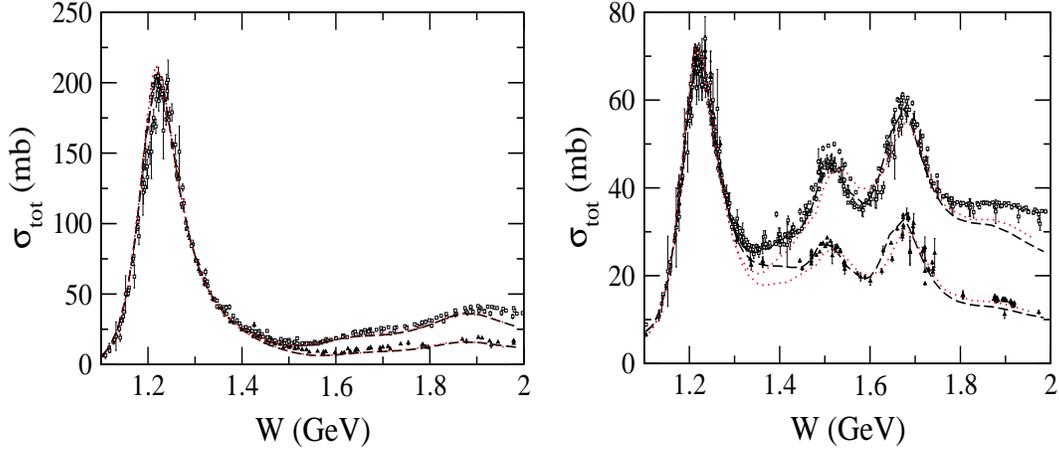}}
\end{center}
\caption{Comparisons between the results from Ref.~\cite{Julia07} 
(dashed curves) and
from the model $B$ (dotted curves) for $\pi N \to X,\pi N$ processes.
Left panel: Predicted total cross-section for the reactions
$\pi^+ p \to X$ (upper set) and $\pi^+ p \to \pi^+ p$ (lower set). 
Right panel: 
predicted total cross-section for the reactions 
$\pi^- p \to X$ (upper set)
$\pi^- p \to \pi^- p + \pi^\circ n$ (lower set); 
Data in both panels are from Refs.~\cite{PDG,SAID}.
}
\protect\label{fig:pin}
\end{figure}

To end this Section, we wish to emphasize that the results of 
Ref.~\cite{Julia07} are extended to the process $\pi^- p \to \eta n$ 
and two models are obtained, reproducing equally well the general 
features of a heterogeneous data-base. This new set of parameters, 
particularly relevant to the investigated process, does not spoil 
the excellent results obtained in Ref.~\cite{Julia07}, and devoted 
mainly to the $\pi N \to \pi N$ observables. In order to illustrate
this latter point, results from Ref.~\cite{Julia07} and our model $B$
are shown in Fig.~\ref{fig:pin}, where Figs. [13] and [14] of 
Ref.~\cite{Julia07} have been complemented with the predictions of 
the model $B$. For the processes with $\pi^+ p$ initial state (left panel
in Fig.~\ref{fig:pin}), results from the two models overlap with each other.
For reactions involving $\pi^- p$ initial states (right panel
in Fig.~\ref{fig:pin}), model $B$ gives deeper minima around $W \approx$ 1.4 GeV
than those reported in Ref.~\cite{Julia07}, with the largest discrepancy
between the two curves being less than 20\%. Comparing the partial-wave 
amplitudes of those models, the main differences appear in the $P_{11}$-
and $P_{13}$-waves.
\subsection{Main features of the $\pi^{-}p \to \eta n$ reaction mechanism}
\label{rm}
In order to shed insights to the main ingredients of the reaction 
mechanism, we concentrate on model $B$. Starting from that model, 
and without further minimizations, we have checked the variations 
of the $\chi^2$ by switching off the nine resonances one by one. 
The results are reported in Table~\ref{tab:Roff}.
%
\begin{table}[ht!]
\begin{tabular}{lccccccccc}
\hline
\hline
         & $S_{11}(1535)$ & $S_{11}(1650)$ & $P_{11}(1440)$ & $P_{11}(1710)$ & $P_{13}(1720)$ 
         & $D_{13}(1520)$ & $D_{13}(1700)$ & $D_{15}(1675)$ & $F_{15}(1680)$ \\
\hline
$\chi^2$ & 48.86 & 2.62 & 3.55 & 2.37 & 2.77 & 2.23 & 1.93 & 2.10 & 2.47 \\
\hline
\hline
\end{tabular}
\caption{Reduced $\chi^2$ per data point for model $B$ with
one resonance switched off (the reduced $\chi^2$ for the full model $B$ being 1.94).}
\protect\label{tab:Roff}
\end{table}
%
%
\begin{figure}[hb!]
\vspace{25pt}
\begin{center}
\mbox{\epsfig{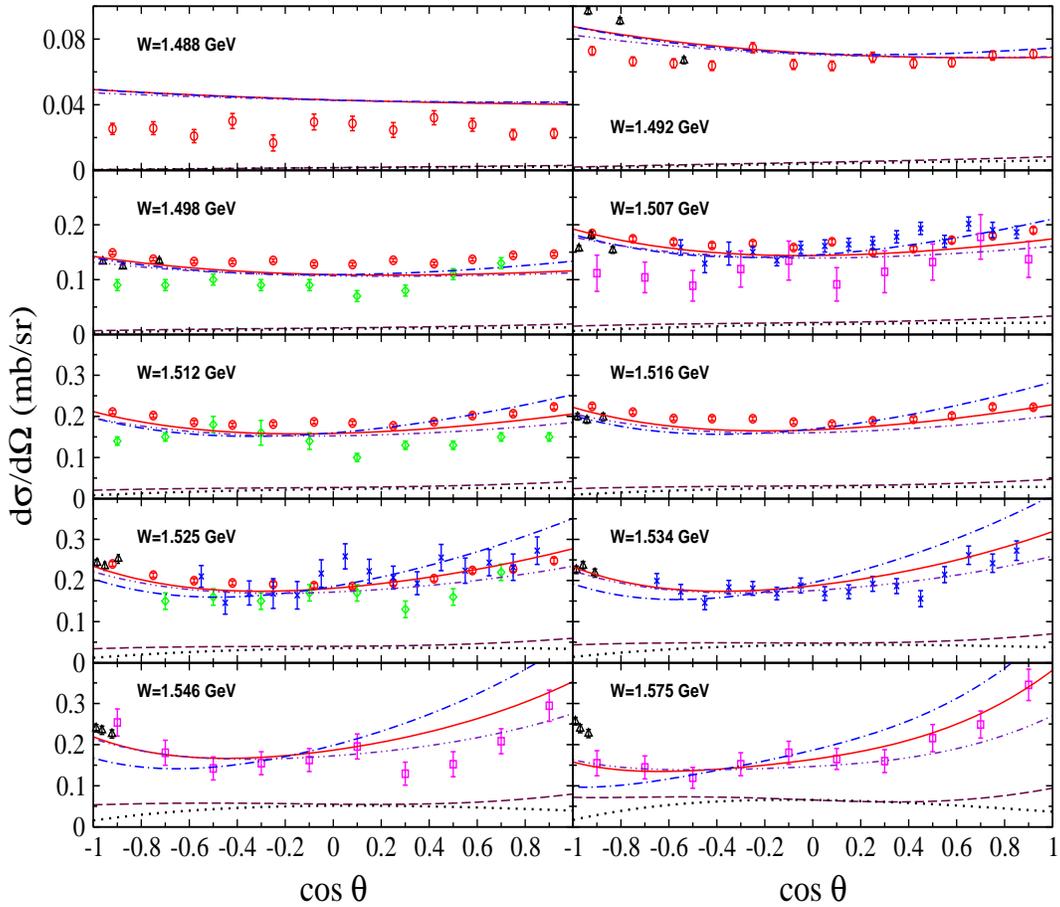}}
\end{center}
\caption{Differential cross-section for the reaction
$\pi^- p \to \eta n$. The full curves correspond to model $B$
and the dotted ones to the non-resonant terms contributions.
The other curves have been obtained by removing one resonance
from that model; the removed resonances are:  
$S_{11}(1535)$ (dashed), 
$P_{11}(1440)$ (dash-dotted), and
$P_{13}(1720)$ (dash-dot-dotted).
Data are as in Fig.~\ref{fig:ds-1}.}
\protect\label{fig:ds-3}
\end{figure}

%
%
\begin{figure}[ht!]
\vspace{25pt}
\begin{center}
\mbox{\epsfig{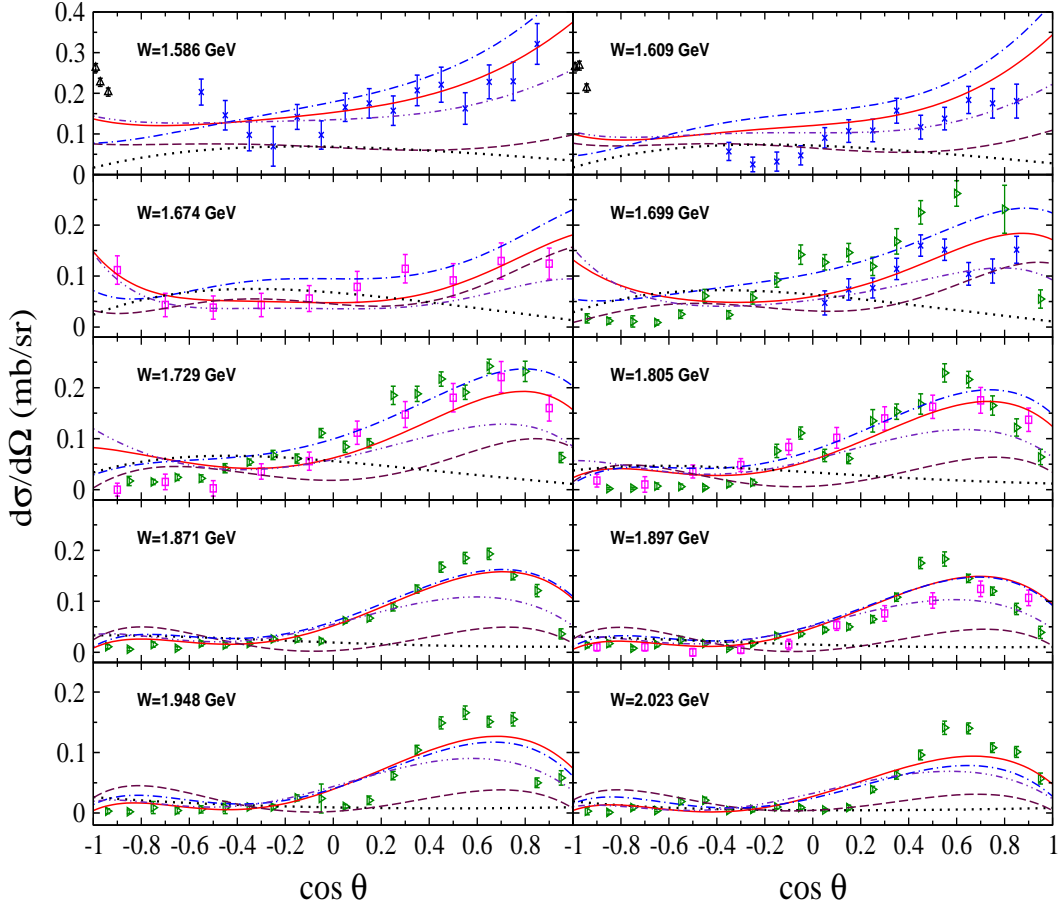}}
\end{center}
\caption{Differential cross-section for the reaction
$\pi^- p \to \eta n$. The curves are is in Fig.~\ref{fig:ds-3}.
Data are as in Fig.~\ref{fig:ds-2}.}
\protect\label{fig:ds-4}
\end{figure}

As expected, the process 
is dominated by the $S_{11}(1535)$ resonance. There are however two 
other resonances playing non-negligible roles, namely, $P_{11}(1440)$ 
and $P_{13}(1720)$. Figures ~\ref{fig:ds-3} and \ref{fig:ds-4} show 
that the importance of those resonances depends on both angle and energy.

The $S_{11}(1535)$ resonance produces more than 80\% of the cross-section 
for the Prakhov~{\it et al.}~\cite{Pra} data. Its importance decreases 
with energy, especially at backward angles, without vanishing out. The 
effect of the $P_{11}(1440)$ resonance becomes visible roughly in the 
energy range 1.525~$\le~W~\le$~1.8 GeV, with a destructive behavior at 
most forward angles. Finally, the $P_{13}(1720)$ appears, in the forward 
hemisphere, around $W~\approx$~1.6 GeV, with the highest contributions at 
$W~\approx$~1.73 GeV and its effect remains constructive.

Although the effect of the $D_{13}(1520)$ on the $\chi^2$ is small, 
it is required to produce the right curvature of the curves at 
low energies.

In conclusion, model $B$ turns out to describe in a satisfactory manner the data set
and embodies a simple reaction mechanism. In the following Section we use hence
that model for further investigations of the $\eta N$ system.

\section{Predictions for the $\eta N$ scattering length and the $\eta p \to \eta p$ 
total cross-section}

The $\eta N$ scattering amplitude in terms of the t-matrix is given by the following relation:

\begin{eqnarray}
f(k) = - \pi \frac{\sqrt {k^2+m^2_N} \sqrt {k^2+m^2_\eta}}{\sqrt {k^2+m^2_N} + \sqrt {k^2+m^2_\eta}}
t_{\eta N}(k,k). 
\label{eq:f}
\end{eqnarray}
Then, the scattering length reads,
\begin{eqnarray}
a_{\eta N} = lim_{k \to 0} f(k) . 
\label{eq:a1}
\end{eqnarray}

\begin{figure}[!ht]
\vspace{35pt}
\begin{center}
\mbox{\epsfig{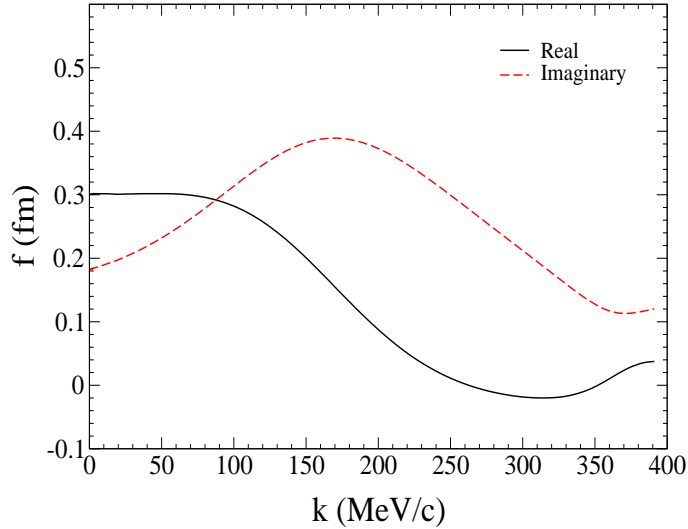}}
\end{center}
\caption{$\eta N$ scattering amplitude $f(k)$ as a function of c.m. momentum, 
within model $B$.
}
\protect\label{fig:RI}
\end{figure}

Fig.~\ref{fig:RI} shows the real and imaginary parts of the function $f(k)$, for model $B$, 
and leads to the following value for the scattering length:
\begin{eqnarray}
a_{\eta N} = (0.30 +i 0.18)~fm. 
\label{eq:a2}
\end{eqnarray}
%

%
%

Efforts since about two decades to determine the ${\eta N}$ scattering length have recently been 
reviewed by several authors~\cite{Sibirtsev02,Green05,Arndt05}. 
A lower limit on the imaginary part, derived 
from the optical theorem and taking into account the recent 
data~\cite{Mor} leads~\cite{Arndt05} to 0.172$\pm$0.009 fm. 
Combining results quoted in those review papers, 
the present knowledge of the imaginary part is:
\begin{eqnarray}
0.17~\lesssim~{\mathcal Im}~a_{\eta N}~\lesssim~0.49~fm, 
\label{eq:a3}
\end{eqnarray}
and our value comes out to be within that range.

For the real part of the scattering length the estimates in the literature give~\cite{Green05},
\begin{eqnarray}
0.27~\lesssim~{\mathcal Re}~a_{\eta N}~\lesssim~ 1.0~fm. 
\label{eq:a4}
\end{eqnarray}
The value extracted in the present work, within model $B$, is close to the lower 
limit. Our value is compatible with those obtained {\it via} chiral effective 
Lagrangians~\cite{Caro00}, most recent solution (G380) from
energy-dependent partial-wave analysis~\cite{Arndt05} of elastic $\pi ^{\pm}p$, 
$\pi ^- p \to \pi^\circ n$, and $\pi ^- p \to \eta n$ scattering data, as well as with older
findings~\cite{Older,Gris00}. Investigations based on chiral perturbation 
approaches~\cite{Kais97,Liu07}
lead to smaller values around 0.2~fm. Finally, coupled-channels calculation within
T-matrix~\cite{Batinic95a,Batinic95b,T-mat}
or
K-matrix~\cite{K-mat}
produce larger values, $0.5~\lesssim~{\mathcal Re}~a_{\eta N}~\lesssim~ 1.0$~fm.

Besides the process $\pi N \to \eta N$, the $\eta$ production using proton or deuteron
beams have also
been investigated using various sets of $\eta N$ scattering lengths reported in the literature, 
as summarized below:

a) $pn \to \eta d$ near threshold data~\cite{pn-data} has been studied within a 
two-step model~\cite{Gris00}, embodying meson-exchange and final-state $\eta N$ 
interactions, and favors small scattering length:
$a_{\eta N}$ = 0.29 + i0.36 fm. 
A microscopic three-body approach in its non-relativistic version~\cite{Garc02} 
reached the conclusion that the data are well reproduced using the results
from Ref.~\cite{Kais97}, $a_{\eta N}$ = 0.42 + i0.34 fm. The relativistic
version of that approach~\cite{Garc05} shows the importance of initial- and
final-state treatments, emphasized also by J\"{u}lich Group~\cite{Baru03}, 
leading to a reduced selectivity on the sets used for the scattering length.

b) $pp \to p p \eta$ ~\cite{Mosk04} and $pn \to p n \eta$~\cite{pn-data} data, 
as well as the above mentioned data have recently been studied within
an effective Lagrangian model~\cite{Shya07} resulting in a reasonable account
of data for $a_{\eta N}$ = 0.51 + i0.26 fm.

c) $\eta$ production in proton-deuteron collisions are being 
studied~\cite{pd}, but at the present time those investigations do not 
allow refinements in determining the $\eta N$ scattering length.

%
\begin{figure}[tbh]
\vspace{35pt}
\begin{center}
\mbox{\epsfig{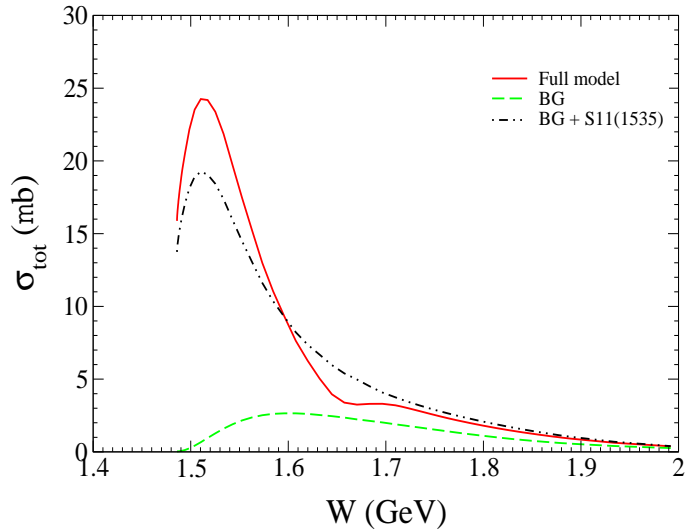}}
\end{center}
\caption{Total cross-section for the reaction
$\eta p \to \eta p$ as a function of total c.m. energy, within model $B$.
}
\protect\label{fig:tot2}
\end{figure}

Findings of various approaches with respect to the $\eta N$ scattering length, 
summarized above, lead then to the ranges for real and imaginary parts as 
reported in Eqs.~(\ref{eq:a3}) and (\ref{eq:a4}). Our value for the real
part being close to the lower bound, excludes the existence of bound
$\eta$-nucleus states.

To end this Section, we show our prediction for the $\eta N$ elastic scattering
total cross-section (Fig.~\ref{fig:tot2}). The background contributions (dashed curve) 
turns out to be small and smoothly varying. 
This latter contribution completed by that of the $S_{11}(1535)$ (dash-dot-dotted curve)
account for a significant portion of the total cross-section predicted by the
full model $B$ (full curve).

%
%

\section{Summary and conclusions}
\label{sum}

A dynamical coupled-channels formalism is used to study the the process
$\pi^{-}p \to \eta n$ in the total center-of-mass energy range
$W~\lesssim$ 2 GeV. The formalism  
embodies, besides non-resonant terms, five intermediate meson-nucleon states,
namely, $\pi N$, $\eta N$, $\pi \Delta$, $\sigma N$, and $\rho N$.

Within this phenomenological approach, 34 adjustable parameters are
introduced, two of them for the non-resonant mechanisms and the others
for the nine nucleon resonances retained in the model search, namely,
$S_{11}(1535)$, $S_{11}(1650)$, $P_{11}(1440)$, $P_{11}(1710)$, $P_{13}(1720)$,
$D_{13}(1520)$, $D_{13}(1700)$, $D_{15}(1675)$, and $F_{15}(1680)$.
That set of resonances corresponds to all known 3 and 4 star resonances
relevant to the energy range investigated here.

In order to determine the parameters and build a model, a data set including
294 measured differential cross-sections, coming from five collaborations, are fitted.
The selection of data points allows to suppress the manifestations of inconsistencies
among available data sets. Our model $B$ reproduces satisfactorily the data, with
a reduced $\chi^2$ = 1.94. 
A detailed study of the reaction mechanism within model $B$ allows to establish
a hierarchy in the roles played by nucleon resonances.
Actually, the dominant resonance turns out to be the $S_{11}(1535)$. The other 
resonances affecting the $\chi^2$ by more than 20\% when switched off, are by
decreasing importance: 
$P_{11}(1440)$, $P_{13}(1720)$, $S_{11}(1650)$,$F_{15}(1680)$,
$P_{11}(1710)$, and $D_{13}(1520)$. Contributions from $D_{13}(1700)$ and 
$D_{15}(1675)$ are found to be negligible.

Model $B$ is used to extract the $\eta N$ scattering length, which comes out
to be: $a_{\eta N}$ = (0.30 +i 0.18)~fm. Both the real and imaginary parts of that quantity
are within the ranges determined from other works.

To improve our knowledge on the $\pi^{-}p \to \eta n$, and consequently on
the $\eta N$ system, further measurements including polarized target asymmetry, are
highly desirable. 
Such experimental results are awaited for thanks to present and/or forthcoming 
pion beams in the following labs:
GSI~\cite{GSI}, ITEP~\cite{epecur}, Fermi Lab~\cite{MIPP}, and JPARC~\cite{JPARC}.

Finally, to take advantage of copious $\eta$ electromagnetic production data, 
the obtained model $B$ appears reliable enough for our in progress investigation of the 
$\gamma p \to \eta p$ reaction within a coupled-channels approach
and a constituent quark model~\cite{He08}.
%
%

\section*{Acknowledgements}

We are grateful to Bill Briscoe and Ben Nefkens for 
enlightening discussions on the status of data. 
This work is 
partially supported by Grant No. FIS2005-03142 from MEC (Spain) 
and FEDER and European Hadron Physics Project RII3-CT-2004-506078.
B.J-D acknowledges the support of the Japanese Society for the Promotion 
of Science (JSPS), grant number: PE 07021, and thanks the nuclear 
theory group at Osaka University for their warm hospitality.
The authors thankfully acknowledge the computer resources, 
technical expertise and assistance provided by 
the Barcelona Supercomputing Center - Centro 
Nacional de Supercomputacion (Spain).

%
%
%

\end{document}